      [1999/12/01 v1.4c Il Nuovo Cimento]
\documentclass{cimento}
\usepackage{graphicx}  
\usepackage{epsfig} 
\title{Dark Matter in a One-dimensional Universe}
\author{Costantino Sigismondi\from{ins:x}}
\instlist{\inst{ins:x} ICRA, International Center for Relativistic 
Astrophysics, and University of Rome La Sapienza, Piazzale Aldo Moro 5 00185 Roma, Italy.}
\PACSes{\PACSit{98.80.H}{Cosmology}
\PACSit{95.35.+d}{Dark Matter}
\PACSit{98.65.-r}{Large Scale Structure}}
\begin{document}

\maketitle

\begin{abstract}
A computer code to simulate temporal evolution of overdensities in a one-dimensional 
Universe is presented for didactic purposes. 
The formation of large scale structures in this one-dimensional universe can be studied both in 
matter or radiation dominated eras. Since large scale structures are already observed at $z \geq 7$,
primordial dark matter overdensities $\delta_{DM}$ which are 90 times larger than the observed barionic $\delta_B$ in 
the cosmic microwave background are required at $z\sim 1000$.
This makes possible non-linear gravitational collapse at redshift $z \geq 7$ and the formation of the structures.
Primordial perturbations $\delta_B\sim 10^{-5}$ do not leave the linear regime of growth without the aid of dark matter's potential wells. 
This code is suitable for commercial worksheets like MSExcel, StarOffice, or OpenOffice.
  
\end{abstract}

\section{Introduction}
Computer codes for simulating the evolution of large scale structure have been massively used in cosmology since years 60s \cite{ref:gil}. There exist complex codes to study numerically the evolution of density perturbations in the expanding Universe \cite{ref:bon,ref:zal}. In those codes it is possible to tune several cosmological parameters representing different physical conditions like coupling between barions and photons, dependence of density parameter $\Omega_0$ on the abundance of various components (hot and cold dark matter, barions, cosmological constant).

The code here presented is based on the Friedmann solution (1922) for the Euclidean Universe as expanding background. Moreover the evolution of a density perturbation does not affect the expansion itself.

The limitation to a one-dimensional analysis greatly reduces mathematical complexity while allows to obtain qualitative and quantitative predictions on the density perturbation evolution.

The evolution of density perturbations of amplitude $\delta=\delta \rho /\rho = 10^{-5}$ starting from the decoupling time down to present time is calculated.
The requirement of dark matter existence in order to explain the formation of large scale structures at redshift $z \geq 7$ is shown.

The advantage of one-dimensional model is the easiness of computation even with a worksheet like MSExcel, it is indicated in teaching to high school up to undergraduate students.
The calculation of equations' coefficients is the only remaining difficulty.

\section{Overdensities in the Universe: a brief history}

Friedmann solution of Einstein equations in the matter era (whose state equation is $P=0$) yields $a(t)\sim t^{2/3}$. 

This regime occurs after the so-called equidensity time in which $\rho$ of matter equals $\rho$ of radiation at $z=40.000$ i.e. 7 years after the Big Bang. 

Going backwards in time, between matter era and radiation era, which is reproducible in the computer code changing the expansion rate according to $a(t)\sim t^{1/2}$,  there is Meszaros effect: density perturbations do not grow because free fall time is larger than expansion time, they stay frozen. 

In particular the derivative of $a(t)$ goes to infinity near singularity, therefore in the early stages of Universe's history, gravitational collapse cannot occur because Universe expands more quickly than any other dynamical time. 	

\section{Matter era}

In the matter era we can study the case of perturbations larger than cosmological horizon (the Hubble radius) and the cases of linear growth and non-linear collapse. 

In this one-dimensional model, overdensities $\delta$ are represented as two pointlike masses $m$ separated by a distance $d$ such that $4/3\pi d^3\bar{\rho}=m$, where $\bar{rho}$ is the average density of the Universe at decoupling time $z=1000$.
The volume of the sphere of radius $d$ contains the mass $m$ -considered pointlike- at the average density of the Universe at the decoupling time. 

In order to simulate an overdensity of $\delta=10^{-5}$ it is sufficient to multiply one of the masses $m$ by ($1+10^{-5}$).

\section{Linear and nonlinear growth of overdensities}

Whatever is the size of the overdensity, it grows according to the approximate law 

$log(\delta(t))\sim log(a(t))\Longrightarrow \delta (t) \propto a(t)$ \cite{ref:pee,ref:pad}. 

We show  this behaviour from globular cluster size ($10^6 M_{\odot}$) to superclusters ($10^{18} M_{\odot}$), in 
figure 1. 

Nonlinear growth before $z\approx7$ is necessary for the onset of large scale structures. In figure 2 dark matter overdensities 90 times larger than the observed at decoupling time in cosmic background radiation (CBR) are 
capable to collapse before present time, regardless on their initial mass or size.

\begin{figure}
\centerline{\epsfxsize=4.9in\epsfbox{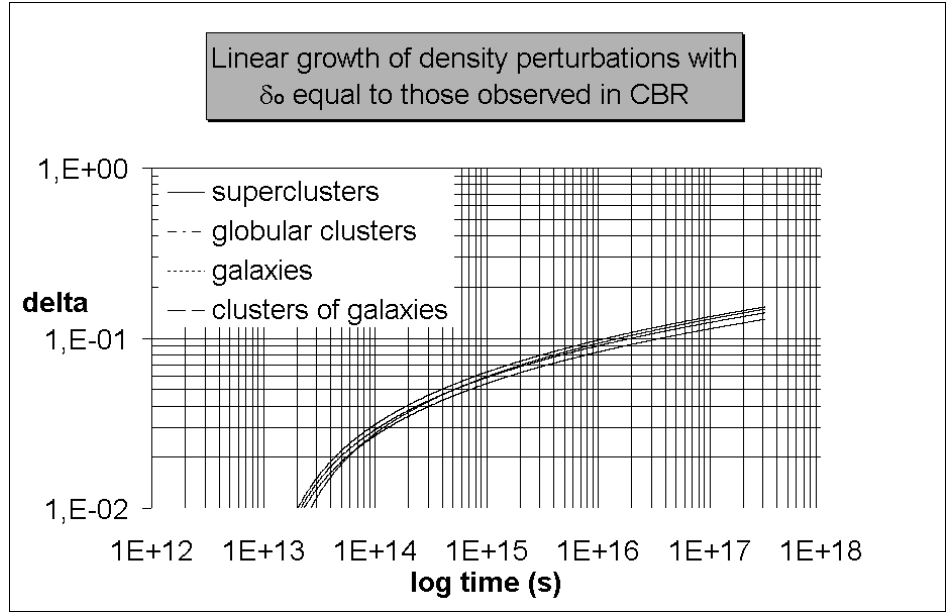}}
\caption{Linear growth of density perturbations with 
$\delta_0=10^{-5}$ equal to those observed in CBR.}
\end{figure}

\begin{figure}
\centerline{\epsfxsize=4.9in\epsfbox{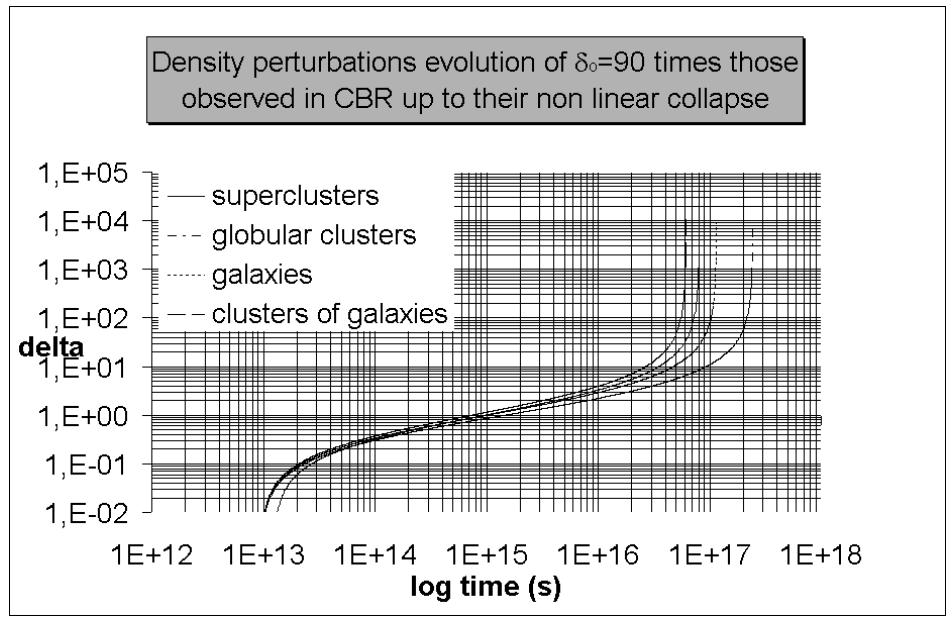}}
\caption{Non-linear growth of perturbations: density perturbations evolution of $\delta_0=90$ times those observed in CBR up to their non linear collapse.}
\end{figure}

\section{Computer code}

In order to reduce the number if iteractions, and at the same time, maintain enough temporal resolution at early stages of Universe's history, we adopted an exponential stepping. 
$t_i \propto e^{(h \cdot i)}$. 
This exponential time is used for computing the discrete expansion factor $\vec{a}(t_1 \ldots t_N)$ according to the opportune dynamical regime.
Once ready the vector $\vec{a}(t)$, it is used in the calculation of the gravitational force between the two masses, at the end of each iteraction, in order to simulate the cosmological expansion.
This code has been implemented in a MSExcel worksheet.

\begin{figure}
\centerline{\epsfxsize=4.4in\epsfbox{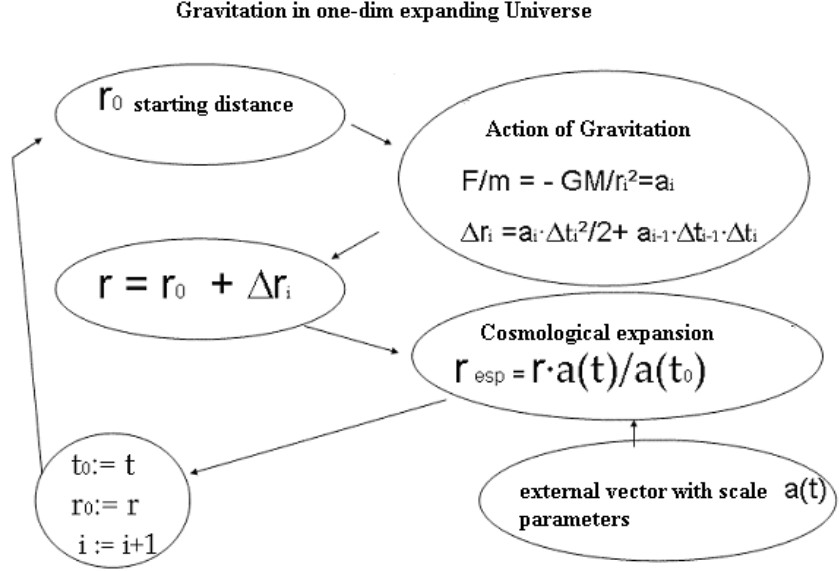}}
\caption{Scheme of one-dimensional Universe's computer code}
\end{figure}

\acknowledgments
This work has been developped at Technical Industrial Institute``Giuseppe Armellini" of Rome during the course ``Leggere il Cielo" -2003 edition- held by the author under
the auspices of the Italian Ministere of Education.

\end{document}